\documentclass[aps,prb,twocolumn,superscriptaddress]{revtex4-1}

\usepackage{graphicx}

\begin{document}

\title{Fully Gapped s-wave-like Superconducting State and Electronic Structures in the Ir$_{0.95}$Pd$_{0.05}$Te$_{2}$ Single Crystals with Strong Spin-orbital Coupling }

\author{D. J. Yu}
\author{F. Yang}
\author{Lin Miao}
\author{C. Q. Han}
\author{Meng-Yu Yao}
\author{Fengfeng Zhu}
\author{Y. R. Song}
\author{K. F. Zhang}
\author{J. F. Ge}
\author{X. Yao}
\affiliation{Key Laboratory of Artificial Structures and Quantum Control (Ministry of Education), Department of Physics and Astronomy, Shanghai Jiao Tong University, Shanghai 200240, China}
\author{Z. Q. Zou}
\affiliation{Center for Analysis and Testing, Shanghai Jiao Tong University, Shanghai 200240, China}
\author{Z. J. Li}
\author{B. Gao}
\affiliation{Shanghai Institute of Microsystem and Information Technology, Chinese Academy of Sciences, Shanghai 200050, China}
\author{D. D. Guan}
\author{Canhua Liu}
\author{C. L. Gao}
\email{clgao@sjtu.edu.cn}
\author{Dong Qian}
\email{dqian@sjtu.edu.cn}
\author{Jin-feng Jia}
\affiliation{Key Laboratory of Artificial Structures and Quantum Control (Ministry of Education), Department of Physics and Astronomy, Shanghai Jiao Tong University, Shanghai 200240, China}

\date{\today}

\begin{abstract}
 Due to the large spin-orbital coupling in the layered 5d-transition metal chalcogenides compound, the occurrence of superconductivity in doped Ir$_{2-x}$Pd$_x$Te$_{2}$ offers a good chance to search for possible topological superconducting states in this system. We did comprehensive studies on the superconducting properties and electronic structures of single crystalline Ir$_{0.95}$Pd$_{0.05}$Te$_{2}$ samples. The superconducting gap size, critical fields and coherence length along different directions were experimentally determined. Macroscopic bulk measurements and microscopic low temperature scanning tunneling spectroscopy results suggest that Ir$_{0.95}$Pd$_{0.05}$Te$_{2}$ possesses a BCS-like s-wave state. No sign of zero bias conductance peak were found in the vortex core at 0.4K.
\end{abstract}


\maketitle

As a new quantum state of matter, topological insulators (TIs) were theoretically proposed involving band inversion due to strong spin-orbital coupling (SOC) and experimentally discovered in compounds with high-Z elements such as HgTe, Bi-based compounds, Sb$_2$Te$_3$\cite{Hasan,Shoucheng,HgTe,BiSb,BiSe,BiTe,SbTe} and so on. Very soon after the discovery of TIs, extensive studies have being carried out in topological matters and topological phenomenon. One of the important topological states is so called topological superconductors (TSCs)\cite{Shoucheng,Schnyder,Qi}, in which the zero-energy mode of Majorana Fermion that is proposed to be useful in topological quantum computation may harbor\cite{Fuliang}. Previously, the most expected TSC is unconventional p-wave superconductor Sr$_2$RuO$_4$\cite{SrRuO1,SrRuO2}. After the conception of topological insulators, many effects for searching the possible TSCs have been put on the superconductors with large spin-orbital coupling. The possible candidates include Cu-intercalated topological insulator Cu$_x$Bi$_2$Se$_3$\cite{Fuliang2,AndrewPRB}, In-doped SnTe\cite{AndoSnTe} and so on. Some signs of zero energy conductance peaks that maybe related to Majorana Fermion were observed \cite{AndoCuBiSe,AndoSnTe,MFScience}. Recently, superconductivity was realized in a layered chalcogenide with high-Z elements: IrTe$_2$ with Pd, Pt and Cu\cite{PdIrTe, PtIrTe, CuIrTe,Lishiyan} substitution or intercalations. Because of the large SOC in this material, it becomes a possible candidate of TSC\cite{PdIrTe}. Though IrTe$_2$ is a layered compound, it is different from typical layered transitional metal dichalcogenides, IrTe$_2$ layers are bonded to each other by significant Te-Te bonding rather than weak Van der Waals force. IrTe$_2$ undergoes a structural phase transition from trigonal phase to triclinic phase at temperature of $\sim$ 270 K\cite{structure,PdIrTe}. The origin of structure transition is still under debate. Fermi surface (FS) nesting, the Rice-Scott ¡°saddle-point¡± mechanism, orbital-induced Peierls instability, crystal field effects, the interlayer hybridization states, the local bonding instability and the anionic depolymerization transition have been proposed\cite{Ding}. With intercalation or substitution at the Ir sites of some nonmagentic elements, the structural transition can be suppressed and bulk superconductivity is induced with T$_c$ up to $\sim$3 K. It'll be very interesting to know the details of superconducting state and how the electronic structures evolve in this system. In this letter, the superconducting properties and electronic structures in the Ir$_{0.95}$Pd$_{0.05}$Te$_{2}$ single crystals were studied by means of dc magnetic susceptibility, electrical resistance, band structure measurements and scanning tunnelling microscopy and spectroscopy (STM/STS) measurements. The superconducting gap size, coherence length, vortex states and band structures were experimental determined directly. We find that Ir$_{0.95}$Pd$_{0.05}$Te$_{2}$ has a fully gapped weak-coupling BCS s-wave-like superconducting state and nearly identical low energy band structure as parent compound IrTe$_2$. No zero bias conductance peak was detected in the vortex core, so topological superconducting state can be rule out in this system at 0.4K.

\begin{figure}
\centering
\includegraphics[width=7.5cm]{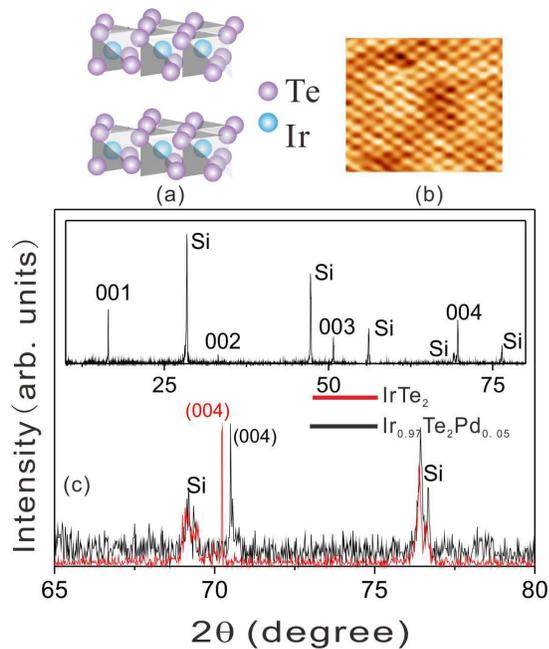}
\caption{(a)Sketch of the IrTe$_2$ crystal structure. (b)Atomic resolved STM image on the cleaving surface of Ir$ _{0.95}$Pd$_{0.05}$Te$_2$ (V=1.5 V, I=190 pA). (c)XRD spectra of the single crystal Ir$ _{0.95}$Pd$_{0.05}$Te$_2$ and IrTe$_2$ at (004) peak. The insert are the large range XRD spectra.}
\end{figure}

High quality single crystals of Ir$_{0.95}$Pd$_{0.05}$Te$_{2}$  were grown using the chemical vapor transport method with iodine as a carrier agent. Doped samples were grown in a single step process in which iridium(99.99\%), tellurium(99.999\%), palladium(99.95\%), and iodine powders (99.99\%) according to the stoichiometric ratio were sealed in an evacuated fused silica ampoule. They were grown with a temperature gradient of 60$\,^{\circ}\mathrm{C}$ with the hot end held at 880$\,^{\circ}\mathrm{C}$. The sample size is about 1mm$\times$1mm$\times$0.2mm. As a comparison, parent compound IrTe$_2$ single crystals were grown out of Te flux using the same method as the previous works\cite{NLWang, PtIrTe}. The crystal structure was checked by x-ray diffraction (XRD) (Bruker) with Cu K$_{\alpha}$ line. The angle-resolved photoemission spectroscopy (ARPES) measurements were performed using 40 - 120 eV photons at Advanced Light Source beamlines 4.0.3 using Scienta R4000 analyzers with base pressures better than 5$\times$10$^{-11}$ torr. Energy resolution is better than 15 meV and angular resolution is better than 0.02 \AA$^{-1}$. Low temperature STM/STS experiments were carried out in ultrahigh vacuum with a base pressure better than 1$\times$10$^{-10}$ Torr. Tungsten tips are used for STM/STS measurements. The dI/dV data of superconducting gaps were obtained via lock-in technique with modulation signal voltage 0.05 mV with a frequency of 985 Hz. The samples are cleaved in situ at 30K as well as at room temperature. The dc magnetization measurements were performed on a Quantum Design Physical Property Measurement System (PPMS). Temperature and magnetic fields dependent resistance measurements were carried out in the PPMS using the standard four-point probe technique with silver paste used for the contacts.

Figure 1 shows the crystal structure of IrTe$_2$ and the XRD spectra from the doped and the undoped crystals. Seen from Fig. 1(c), except the Si reference peak, all the diffraction peaks can be indexed by $(00l)$ peaks according to P$\bar{3}$m1 structure from Ir$_{1-x}$Pd$_{x}$Te$_{2}$ with FWHM less then 0.05 degree which indicates the high crystalline quality of the samples. Si reference peaks were used to correct the system error. The (004) peaks were used to calculate the c-axis lattice constant. Based on Bragg condition, the calculated c-axis lattice constant is 5.386 \AA\ and 5.372 \AA\ for IrTe$_2$ and Ir$_{1.95}$Pb$_{0.05}$Te$_2$, respectively. After x=0.05 Pd doping, the lattice contracts by $\sim$ 0.26\% that is consistent with the Pd-Te substitution known from the powder samples\cite{PdIrTe}. After cleaving, the surface is Te-terminated. Fig. 1(b) shows well-ordered hexagonal lattice on the surface.

\begin{figure}

\centering
\includegraphics[width=7.5cm]{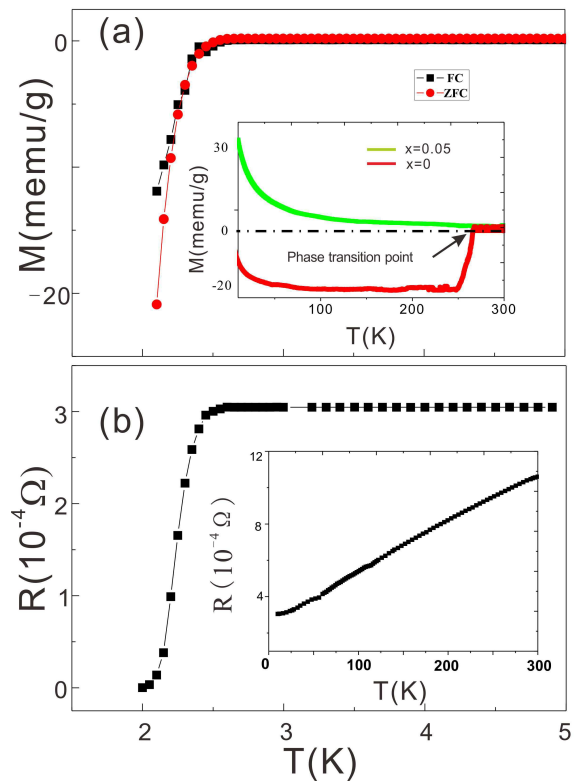}
\caption{(a) Temperature dependence of the dc magnetization under 20 Gauss field of $\rm{Ir _{0.95}Pd_{0.05}Te_{2}}$ with zero field cooling (ZFC) and field cooled (FC). Diamagnetic signal is observed at $\sim$ 2.5K. Insert: Temperature dependence of of the dc magnetization of $\rm{Ir _{0.95}Pd_{0.05}Te_{2}}$ and IrTe$_2$ with 2T magnetic field applied perpendicular to the ab plane. (b) Temperature dependence of the resistance of $\rm{Ir _{0.95}Pd_{0.05}Te_{2}}$ crystals with current flowing in the ab plane. Superconducting onset temperature is $\sim$ 2.5K. Insert: Temperature dependence of the resistance for $\rm{Ir _{0.95}Pd_{0.05}Te_{2}}$ with 2T magnetic field applied perpendicular to the ab plane.}
\end{figure}

Typical low-temperature dc magnetic signal and electrical resistance of Ir$_{0.95}$Pd$_{0.05}$Te$_{2}$ samples are shown in Fig. 2. A diamagnetic behavior, characteristic of superconducting state, is observed below $\sim$2.5K. The magnetization signal obtained at 2.1 K is about 60$\% $ of that expected for full diamagnetism. This represents a conservative lower limit to the true superconducting volume fraction because the diamagnetic magnetization is still decreasing steeply at the temperature where the field is applied for the zero field cooling measurement. Figure 2b shows the temperature dependence of the resistance of $\rm{Ir _{0.95}Pd_{0.05}Te_{2}}$ , measured in the ab plane. Consistent with magnetic measurement, resistance of the sample decreases at $\sim$ 2.5K and reaches zero at $\sim$ 2K, suggesting bulk superconductivity. As presented in the insert of Fig. 2(a), with x=0.05 Pd substitution the magnetic signal increases monotonically in the doped samples with the decrease of temperature under the magnetic filed of 2T. The structure phase transition is completely suppressed in Ir$_{0.95}$Pd$_{0.05}$Te$_2$, while IrTe$_2$ shows a sharp transition at $\sim$ 270K\cite{PdIrTe}, marked by black arrow.

\begin{figure}
\centering
\includegraphics[width=8cm]{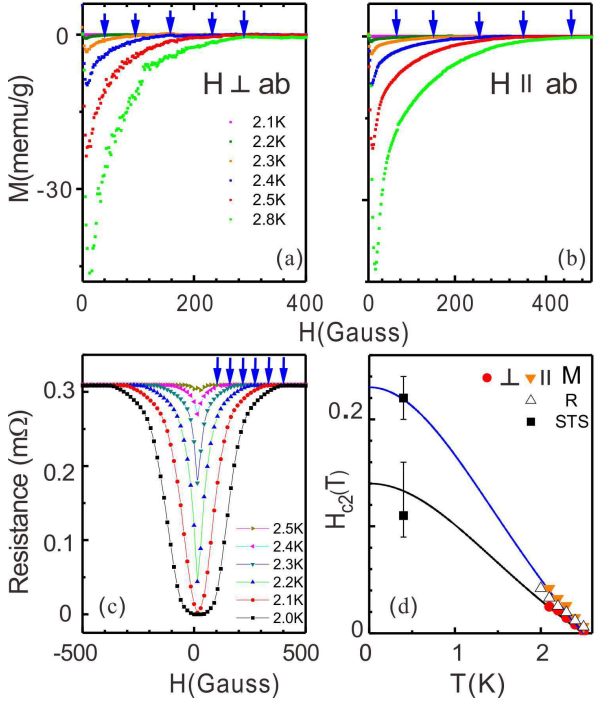}
\caption{M(H) curves at T = 2.1 K, 2.2 K, 2.3 K,2.4 K, 2.5 K and 2.8 K with magnetic field (a) perpendicular to ab plane. (b) parallel to ab plane. (c) Resistance as a function of magnetic field applied in ab plane at T = 2.0 K, 2.1 K, 2.2 K, 2.3 K,2.4 K and 2.5 K. Blue arrows marking the position of $\rm{H_{c2}(T)}$ . (d) $\rm{H_{c2}(T)}$ determined from $M(H)$, R(H) and STS data at different temperatures. Solid curves are fitting curves to the magnetic signal measurements.}
\end{figure}

\begin{figure}
\centering
\includegraphics[width=8.0cm]{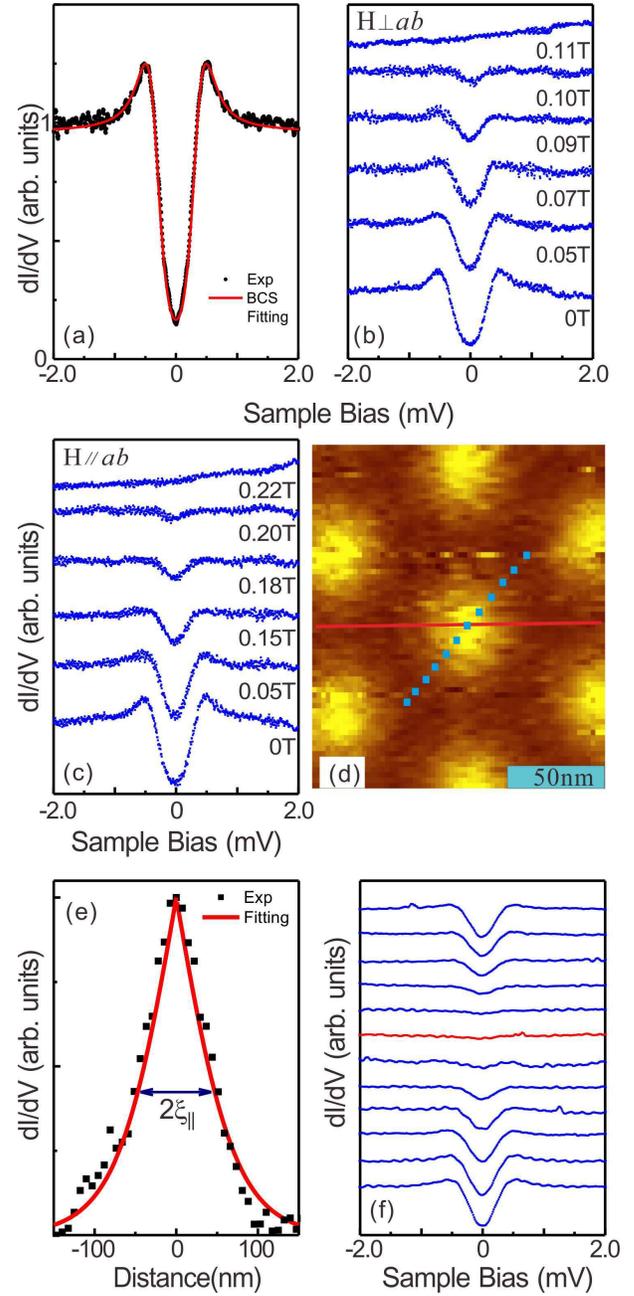}
\caption{(a)Typical superconducting gap (dI/dV spectra) measured at 0.4K. The distance between the tip and sample is set to be $\sim$ 1nm. Superconducting gap is further confirmed by checking the evolution of the dI/dV spectra with magnetic fields(b) perpendicular to ab and (c) parallel to ab plane. (d)Superconducting vortex lattice measured by STS mapping. Applied magnetic field is 0.05 T. Red curve shows how to get the vortex's line profile. (e)Line profile of the single vortex. Red curve is the fitting curve to extract the coherence length. (f)dI/dV curves measured at different position according to blue dots in (d).}
\end{figure}

Detailed measurements of the magnetic signals and the resistance as a function of magnetic fields are presented in Fig. 3. Fig. 3(a) and (b) show the magnetization curves at different temperature (from 2.1K  to  2.8K) along out-of-plane($\perp)$ and in-plane($\parallel$) direction, respectively. Seen from the magnetization curves, $\rm{Ir _{0.95}Pd_{0.05}Te_{2}}$ has typical type-II superconductor behavior. As shown in the Fig. 3(a,b), the blue arrows mark the position of $\rm{H_{c2\perp}}$ and $\rm{H_{c2\parallel}}$ where diamagnetic signals disappear at different temperature. In the resistance data (Fig. 3c), the upper critical field $\rm{H_{c2}}$ was defined at the superconducting onset temperature. Fig. 3(d) summaries the H$_{c2}$ as a function of temperature. The linear temperature dependence close to $\rm{T_{c}}$ is obtained for $\rm{H_{c2}}$, suggesting the dominance of only one type of bulk carrier, which is consistent with ARPES measurement shown below. The solid line is the fitting to Werthamer-Helfand-Hohenberg (WHH) theory based on the magnetic measurements. We obtained the upper critical field at zero temperature $\rm{H_{c2,\perp}(0)}$=0.16$\pm$0.05 T and $\rm{H_{c2\parallel}(0)}$ =0.33 $\pm$0.05 T. From $\rm{H_{c2\perp}}$, the coherence length  $ \xi_{\parallel}=\sqrt{\Phi_0/2\pi \rm{H_{c2\perp}}  }=45nm$ is obtained, while from $\rm{H_{c2\parallel}}$ we use $ \xi_{\parallel}\xi_{\perp}=\Phi_0/2\pi \rm{H_{c2\parallel}}$ and obtain $ \xi_{\perp}= 22 nm$. So the dimensionless anisotropy parameter $\gamma=\xi_{\parallel}/\xi_{\perp}\sim$ 2, which is smaller than another layered superconductor NbSe$_2$ ($\gamma\sim$3).

\begin{figure}[b]
\centering
\includegraphics[width=8.5cm]{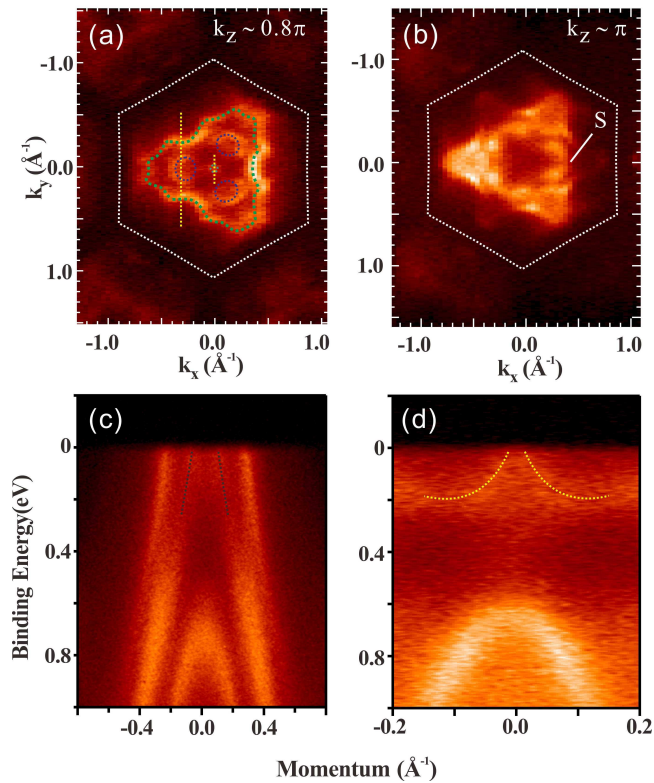}
\caption{(a)Fermi surface at k$_z$ $\sim$ 0.8$\pi$ on Ir$_{0.95}$Pd$_{0.05}$Te$_2$ measured at 30K. Green and blue dotted curves present the measured Fermi surfac at room temperature on IrTe$_2$. No obvious different was observed between Ir$_{0.95}$Pd$_{0.05}$Te$_2$ and IrTe$_2$. (b)FS of Ir$_{0.95}$Pd$_{0.05}$Te$_2$ at k$_z\sim\pi$. "S" labels the same saddle point as in IrTe$_2$\cite{Ding}. (c) and (d) ARPES spectra at the momentum space as indicated in (a) by yellow dotted line. Black dotted line marks the hole-like bands crossing the Fermi level that sink beneath the Fermi level in IrTe$_2$ at low temperature.}
\end{figure}

The properties of the superconducting state are further explored by low temperature STM/STS. Fig. 4(a) shows the typical STS spectra obtained on the cleaved surface at 0.4K. Well-defined coherence peaks corresponding to the superconducting state was observed clearly. The STS spectra can be nicely fitted using BCS-type s-wave formula, which gives a superconducting gap of 0.36$\pm$0.02 meV at 0.4K. Assuming the gap follows BCS theory, $\Delta(T)$=$\Delta(0)(1-T/T_c)^{1/2}$, we get $\Delta(0)\sim$0.39$\pm$0.02 meV, yielding the BCS ratio 2$\Delta$/k$_B$T$_c$ $\sim$ 3.6. This value indicates a weak-coupling BCS-type superconductor (for weak coupling limit s-wave, BCS ratio $\sim$ 3.5). The STS spectra as a function of applied magnetic fields are shown in Fig. 4(b) and 4(c). With the increase of applied magnetic field, the coherence peak faded away gradually. The energy gap disappears at about 0.11 T and 0.22 T when magnetic filed perpendicular or parallel to ab plane. It is worth noting that these two values are position dependent due to the existence of the vortex (vortices are shown in Fig. 4(d)) and the variation is much smaller when the magnetic field is applied in the ab plane. Nevertheless, the upper critical fields obtained in STS are well consistent with macroscopic bulk measurements as shown in Fig. 3(d). Superconducting vortex was observed by STS mapping. Fig. 4(d) shows the vortex with nice close-packed arrangement. From the size of the vortex, as shown in fig. 4(e), the in plane coherence length $\xi(T=0.4K)$ of $\sim$51$\pm$ 5 nm is obtained from the fitting of the line profile (red line in Fig. 4d) of the vortex, which is nearly consistent with the value from the bulk magnetic measurement. Small BCS ratio, s-wave-like gap function and long coherence length suggest that Ir$ _{0.95}$Pd$_{0.05}$Te$_2$ is a fully gapped and weak coupling BCS-type s-wave superconductor. In addition, we carefully checked the STS at different position of vortex as shown in Fig. 4(f). Red curve in Fig. 4(f) is collected right at the center of a vortex. Outside the vortex, there is superconducting gap and gap closes at the center of the vortex. Under our experimental temperature (0.4K), no sign of zero bias conductance peak was observed at any position of vortex, which means even if there will be non-trivial superconducting states, much lower temperature is needed. From the STS measurement, we think we can rule out the possibility of topological superconducting state in this system at 0.4K.

Finally, the electronic structures of Ir$ _{0.95}$Pd$_{0.05}$Te$_2$ are studied by ARPES. Same as IrTe$_2$, the energy bands of Ir$ _{0.95}$Pd$_{0.05}$Te$_2$ show strong k$_z$ dispersion. By tuning incident photon energy, we can change the detectable
momentum along the normal direction (k$_z$). In-plane FS at different k$_z$ was obtained by integrating the spectra weight with an energy window of 15 meV at Fermi energy. Changing of FS topology was observed in previous studies in IrTe$_2$ below and above the phase transition temperature\cite{JAP,Ding}. Fig. 5(a) shows the FS of Ir$ _{0.95}$Pd$_{0.05}$Te$_2$ measured at 30K overlaid with the room temperature FS (green and blue dotted lines) of IrTe$_2$ from reference [24] at the exactly same k$_z$ ($\sim$ 0.8$\pi$) point using 90eV photon energy. Clearly, after Pd substitution, the FS of superconducting samples at low temperature is the same as that of IrTe$_2$ at room temperature. There are one outer big FS, three small Fermi pockets and a tiny Fermi pocket around zone center. Meanwhile, the three small Fermi pockets (blue dotted lines in Fig. 5-a) disappear at 30K in IrTe$_2$ samples\cite{Ding}. Fig. 5(c) and (d) show energy bands along the direction indicated by the yellow dotted lines in Fig. 5(a). Known from the band mapping, all the FSs are formed by hole-like bands. Since Ir$ _{0.95}$Pd$_{0.05}$Te$_2$ has no phase transition but has identical FS as that of the IrTe$_2$ sample above phase transition temperature, it implies that itinerant origin such as FS nesting or saddle point nesting may not play key roles for the phase transition in this system. Phase transition is more likely related to local phenomena. For example, the depolymerization-polymerization of Te bonds was proposed to be involved in the phase transition\cite{Dem}. And it was also proposed that Ir 5d orbital reconstruction governed the charge and orbital instability in IrTe$_2$\cite{PtIrTe}. In addition, charge modulation originated from the periodic dimerization of Te atoms was observed in previous STM expereiments\cite{Wu}. Fig. 5(b) presents the FS near $\pi$ point. At this k$_z$ point, saddle point (marked as "S" in the figure) observed in IrTe$_2$\cite{Ding} remains in Ir$ _{0.95}$Pd$_{0.05}$Te$_2$, which leave the possibility that the superconducting instability can be related to the van Hove singularity at saddle points\cite{Ding}. This scenario could be studied by ultra-low temperature ARPES experiments in the future.

In summary, we did comprehensive studies of the superconducting properties and electronic structures of Ir$ _{0.95}$Pd$_{0.05}$Te$_2$. By combination of the macroscopic and microscopic measurements, the superconducting gap size, coherence length, electronics structures as well as the vortex states were determined. At 0.4K, the samples present a BCS-type s-wave-like superconducting behavior. Though no exotic superconducting state was found in this system, the hexagonal lattice structure of Ir$ _{0.95}$Pd$_{0.05}$Te$_2$ is very suitable for expitally growth TIs. In our previous work, Bi$_2$Se$_3$ (Bi$_2$Te$_3$) TI films were successfully grown on s-wave superconductor NbSe$_2$\cite{WangMX,Xujinpeng}. This type of heterostructure utilizing the superconducting proximity effect perpendicular to the ab plane, so called vertical geometry. Superconducting Ir$_{1-x}$Pd$_x$Te$_2$ will be another excellent candidate for this geometry since its vertical coherence length is much larger than that of NbSe$_2$ ($\xi \sim$ 3 nm). The heterostructure of TI/Ir$_{1-x}$Pd$_x$Te can be a good candidate for exploring topological superconductor related phenomena in the future.

This work is supported by National Basic Research Program of China (Grants No. 2012CB927401, 2011CB921902, 2013CB921902, 2011CB922200), NSFC (Grants No. 91021002,  10904090, 11174199, 11134008), the SCST, China (Grants No. 12JC1405300, 13QH1401500, 10JC1407100, 10PJ1405700, 11PJ405200). The Advanced Light Source is supported by the Director, Office of Science, Office of Basic Energy Sciences, of the US Department of Energy under Contract DE-AC02-05CH11231. D.Q. acknowledges additional supports from the Top-notch Young Talents Program and the Program for Professor of Special Appointment (Eastern Scholar) at Shanghai Institutions of Higher Learning.


\begin{thebibliography}{50}
\bibitem{Hasan}M. Z. Hasan and C. L. Kane, Rev. Mod. Phys. \textbf{82}, 3045 (2010).
\bibitem{Shoucheng}X.-L. Qi and S.-C. Zhang, Rev. Mod. Phys. \textbf{83}, 1057 (2011).
\bibitem{HgTe}B. A. Bernevig, T. L. Hughes, and S.C. Zhang, Science \textbf{314}, 1757 (2006).
\bibitem{BiSb}D. Hsieh, D. Qian, L. Wray, Y. Xia, Y. S. Hor, R.J. Cava and M.Z. Hasan, Nature \textbf{452}, 970 (2008).
\bibitem{BiSe}Y. Xia, D. Qian, D. Hsieh, L. Wray, A. Pal, H. Lin, A. Bansil, D. Grauer, Y.S. Hor, R.J. Cava, M.Z. Hasan, Nature Phys. \textbf{5}, 398 (2009).
\bibitem{BiTe}Y.L. Chen, J.-H. Chu, J.G. Analytis, Z.K. Liu, K. Igarashi, H.-H. Kuo, X.L. Qi, S.K. Mo, R.G. Moore, D.H. Lu, M. Hashimoto, T. Sasagawa, S.C. Zhang, I.R. Fisher, Z. Hussain, and Z.X. Shen, Science \textbf{325}, 178 (2009).
\bibitem{SbTe}H.J. Zhang, C.X. Liu, X.L. Qi, X. Dai, Z. Fang, and S.C. Zhang, Nat. Phys. \textbf{5}, 438 (2009).

\bibitem{Schnyder}A. Schnyder, S. Ryu, A. Furusaki, A. Ludwig, Phys. Rev. B, \textbf{78}, 195125 (2008).
\bibitem{Qi}X.L. Qi, T.L. Hughes, S. Raghu, S.C. Zhang, Phys. Rev. Lett. \textbf{102}, 187001 (2009).
\bibitem{Fuliang}L. Fu, C. L. Kane, Phys. Rev. Lett. \textbf{100}, 096407 (2008).
\bibitem{SrRuO1}A. P. Mackenzie and Y. Maeno, Rev. Mod. Phys. \textbf{75}, 657 (2003).
\bibitem{SrRuO2}Y. Maeno, S. Kittaka, T. Nomura, S. Yonezawa, K. Ishida, J. Phy. Soc. J. \textbf{81}, 011009 (2012).
\bibitem{Fuliang2}L. Fu and E. Berg, Phys. Rev. Lett. \textbf{105}, 097001 (2010).
\bibitem{AndrewPRB}L.A. Wray, S. Xu, Y. Xia, D. Qian, A.V. Fedorov, H. Lin, A. Bansil, L. Fu, Y.S. Hor, R.J. Cava, M.Z. Hasan, Phys. Rev. B, \textbf{83}, 224516 (2011).

\bibitem{AndoSnTe}S. Sasaki, Z. Ren, A. A. Taskin, K. Segawa, L. Fu, and Y. Ando, Phys. Rev. Lett. \textbf{109}, 217004 (2012).
\bibitem{AndoCuBiSe}S. Sasaki, M. Kriener, K. Segawa, K. Yada, Y. Tanaka, M. Sato, and Y. Ando, Phys. Rev. Lett. \textbf{107}, 217001 (2011).
\bibitem{MFScience}V. Mourik, K. Zuo, S. M. Frolov, S. R. Plissard, E. P. A. M. Bakkers, L. P. Kouwenhoven, Science \textbf{336}, 6084 (2012).
\bibitem{PdIrTe}J.J. Yang, Y. J. Choi, Y. S. Oh, A. Hogan, Y. Horibe, K. Kim, B. I. Min, and S.W. Cheong
Phys. Rev. Lett. \textbf{108}, 116402 (2012).
\bibitem{PtIrTe}D. Ootsuki, Y. Wakisaka, S. Pyon, K. Kudo, M. Nohara, M. Arita, H. Anzai, H. Namatame, M. Taniguchi, N.L. Saini, T. Mizokawa, Rev. B \textbf{86}, 014519 (2012).
\bibitem{CuIrTe}M. Kamitani1, M. S. Bahramy, R. Arita1, S. Seki, T. Arima, Y. Tokura, and S. Ishiwata, Phys. Rev. B \textbf{87}, 180501(R) (2013).
\bibitem{Lishiyan}S.Y. Zhou, X.L. Li, B.Y. Pan, X. Qiu, J. Pan, X.C. Hong, Z. Zhang, A.F. Fang, N.L. Wang, S.Y. Li, EPL, \textbf{104}, 27010 (2013).
\bibitem{structure}H.B. Cao, B.C. Chakoumakos, X. Chen, J.Q. Yan, M.A. McGuire, H. Yang, R. Custelcean, H.D. Zhou, D.J. Singh, D. Mandrus, Phys. Rev. B \textbf{88}, 115122 (2013).
\bibitem{Ding}T. Qian, H. Miao, Z.J. Wang, X. Liu, X. Shi, Y.B. Huang, P. Zhang, N. Xu, P. Richard, M. Shi, M. H.
Upton, J.P. Hill, G. Xu, X. Dai, Z. Fang, H.C. Lei, C. Petrovic, A.F. Fang, N.L. Wang, H. Ding, arXiv:1311.4946 (2013).
\bibitem{NLWang}A.F. Fang, G. Xu, T. Dong, P. Zheng, N. L. Wang, Sci. Rep., \textbf{3}, 1153 (2012).
\bibitem{JAP}D. Ootsuki, S. Pyon, K. Kudo, M. Nohara, M. Horio, T. Yoshida, A. Fujimori, M. Arita, H. Anzai, H. Namatame,
M. Taniguchi, N.L. Saini, T. Mizokawa, J. Phys. Soc. Jpn. \textbf{82} 093704 (2013).
\bibitem{Dem}Y.S. Oh, J. J. Yang, Y. Horibe, and S. W. Cheong, Phys. Rev. Lett. \textbf{110}, 127209 (2013).
\bibitem{Wu}P.J. Hsu, T. Mauerer, M. Vogt, J.J. Yang, Y.S. Oh, S.W. Cheong, M. Bode, W.D. Wu, Phys. Rev. Lett., \textbf{111}, 266401 (2013).
\bibitem{WangMX}M.X. Wang, C.H. Liu, J.P. Xu, F. Yang, L. Miao, M.Y. Yao, C.L. Gao, C. Shen, X.C. Ma, X. Chen, Z.A. Xu, Y. Liu, S.C. Zhang, D. Qian, J.F. Jia, Q.K. Xue, Science \textbf{336}, 52 (2012).
\bibitem{Xujinpeng}J.P. Xu, C.H. Liu, M.X. Wang, J.F. Ge, Z.L. Liu, X.J. Yang, Y. Chen, Y. Liu, Z.A. Xu, C.L. Gao, D. Qian, F.C. Zhang, Q.K. Xue, Jin-Feng Jia, arxiv:1312.3713 (2013).

\end{thebibliography}
\end{document}